# Control of rodent sleeping sickness disease by surface functionalized amorphous nanosilica

Dipankar Seth[1, 3], Mritunjay Mandal[2], Nitai Debnath[1], Ayesha Rahman[1], N. K. Sasmal[2], Sunit Mukhopadhyaya[3], and Arunava Goswami[1]

[1]Biological Sciences Division, Indian Statistical Institute, 203 B.T. Road, Calcutta- 700 108, West Bengal, India.

[2]Department of Veterinary Parasitology and [3]Department of Veterinary Pathology, West Bengal University of Animal and Fisheries Sciences, 37 Kshudiram Bose Sarani, Calcutta 700 037, West Bengal, India.

*To whom correspondence should be addressed. E-mail: agoswami@isical.ac.in (A.G.)

Wild animals, pets, zoo animals and mammals of veterinary importance heavily suffer from trypanosomiasis[1]. Drugs with serious side effects are currently mainstay of therapies used by veterinarians. Trypanosomiasis is caused by *Trypanosoma sp.* leading to sleeping sickness in humans. Surface modified (hydrophobic and lipophilic) amorphous nanoporous silica molecules could be effectively used as therapeutic drug for combating trypanosomiasis. The amorphous nanosilica was developed by top-down approach using volcanic soil derived silica (Advasan; 50-60 nm size with 3-10 nm inner pore size range) and diatomaceous earth (FS; 60-80 nm size with 3-5 nm inner pore size range) as source materials[2-5]. According to WHO and USDA standards amorphous silica has long been used as feed additives for several veterinary industries and considered to be safe for human consumption. The basic mechanism of action of these nanosilica molecules is mediated by the physical absorption of HDL components in the lipophilic nanopores of nanosilica. This reduces the supply of the host derived cholesterol, thus limiting the growth of the *Trypanosoma sp. in vivo*.

Parasites like *Trypanosoma evansi* isolated from naturally infected horses were artificially infected into albino mice. The infection causes 100% mortality in mice within 72 ± 24 hours post infection. Incidentally mice with *T. evansi* have also been used as model systems for studying human trypanosomiasis. Serum derived high-density lipoprotein (HDL) could kill *Trypanosoma sp.* via formation of anion-selective pores in the lysosomal membrane of parasite. Depletion of host HDL is critical for proliferation of *T. evansi*[6-10]. We demonstrate in this research paper that Advasan nanosilica could be effectively used as novel drugs for treatment of rodent trypanosomiasis.

21 day old homozygous albino male mice bred at the mouse facility of West Bengal University of Animal and Fisheries Sciences were artificially infected with *T. evansi* and nanosilica (Advasan and FS) were given as oral dose 1 day post infection. After 72±24 hours post infection, 100% control infected mice died. 100% Advasan nanosilica treated mice survived 192 ± 24 hours (period of observation: 216 hours), whereas FS nanosilica treated mice survived for 120±24 hours. Post-mortems were done for studying gross pathological changes.

| Cholesterol Types | Control mice Serum (mgm / dl) | *T. evansi* Infected mice serum (mgm / dl) | Nanosilica treated *T. evansi* infected mice serum (mgm / dl) | |
| --- | --- | --- | --- | --- |
| | | | FS | Advasan |
| Serum Cholesterol | 289 ± 6.23 | 130 ± 8.03 | 208 ± 5.25 | 203 ± 3.35 |
| **Serum HDL Cholesterol** | **46 ± 5.03** | **29 ± 0.63** | **39 ± 1.13** | **41 ± 1.86** |
| Serum LDL Cholesterol | 194 ± 6.27 | 71 ± 1.67 | 154 ± 3.43 | 134 ± 7.13 |
| Serum VLDL Cholesterol | 49 ± 2.09 | 30 ± 1.04 | 15 ± 1.08 | 28 ± 4.76 |
| Serum Triglycerides | 247 ± 4.08 | 150 ± 5.66 | 78 ± 8.02 | 139 ± 5.47 |

Table 1 legend. The different serum cholesterol components as well as the total content were measured 72 hours post infection. In *T. evansi* infected mice serum, significant decrease in the amount of the HDL was observed. On application of nanosilica (0.025 mg/ ml) HDL levels were found to be near normal levels.

Table 1 shows that during the course of the *T. evansi* infection process all the serum cholesterol components decrease. HDL is the most notable amongst them. But after the administration of the lipophilic nanosilica at the particular dose mentioned, the levels of HDL came back to the normal physiological level. This is essential for the quick recovery of the mice to normal health. The application of higher doses of the nanosilica depletes the cholesterol level completely and impedes the process of recovery of the mice (data not shown). As cholesterol is necessary for a number of normal physiological processes, maintenance of the requisite physiological level has to be taken into consideration while designing the drug dosage.

Table 1 also shows that FS nanosilica treatment could bring the HDL level back to normal, but it also decreased VLDL level in infected mice to a very low level. As VLDL is necessary is necessary for normal body functions, therefore mice died early out of secondary reasons in case of FS nanosilica treatment.

We found that 100% of the infected mice (n=18) treated with nanosilica survived while 100% of the untreated infected mice (n=18) died within 72±24 of inoculation. The parasite burden was reduced by margin of 92±5 % (n=18) compared to untreated infected mice (n=18) after 72±24 hours of infections.

To the best of our knowledge, this scientific report stands as a first scientific report in this area of research worldwide. All this open a new era `metabolomic modulator' based drug development using nanotechnology and in particular, controlling human trypanosoma growth both *in vitro* and *in vivo* might be possible in near future.

================================
**Total word count: 781 (excluding reference)**
**Table: 1**
**Reference: 10 (ref 8 and 9 could be deleted after review)**